\documentclass[conference]{IEEEtran}
\IEEEoverridecommandlockouts
\usepackage{cite}
\usepackage{xurl}
\usepackage{amsmath,amssymb,amsfonts}
\usepackage{algorithmic}
\usepackage{graphicx}
\usepackage{textcomp}
\usepackage{xcolor}
\usepackage{subcaption}
\usepackage[hidelinks]{hyperref}
\usepackage{url}
\usepackage{enumitem}
\usepackage{acronym}
\usepackage{array}
\usepackage{booktabs}
\def\BibTeX{{\rm B\kern-.05em{\sc i\kern-.025em b}\kern-.08em
    T\kern-.1667em\lower.7ex\hbox{E}\kern-.125emX}}

\IEEEoverridecommandlockouts
\IEEEaftertitletext{\vspace{-20pt}} % Adjusts the gap globally between title/authors and abstract
\begin{document}

\title{A Space-Agnostic Visual Game Analytics Tool with Adaptive Spatial Reconstruction for Mixed Reality Game Development
\\

% \thanks{The code and dataset for this study are available at \url{https://github.com/NahianAlindo/multilabel-perceptual-bug-detection}}
}

\author{
\IEEEauthorblockN{
Nahian Rifaat$^{*}$,
Fedir Skliar$^{\dagger}$,
Parisa Sargolzaei$^{*}$,
Loutfouz Zaman$^{*}$
}

\IEEEauthorblockA{
$^{*}$\textit{Ontario Tech University}, Oshawa, Ontario, Canada\\
$^{\dagger}$\textit{Igor Sikorsky Kyiv Polytechnic Institute}, Kyiv, Ukraine\\[2pt]
nahian.rifaat@ontariotechu.net,
fedos.sklyar@gmail.com,
parisa.sargolzaei@ontariotechu.net,
loutfouz.zaman@ontariotechu.ca
}
\vspace{-10pt}
}

\IEEEpubid{\makebox[\columnwidth]{979-8-3315-9476-3/26/\$31.00 ~\copyright2026 IEEE\hfill}
\hspace{\columnsep}\makebox[\columnwidth]{ }}
\maketitle
\IEEEpubidadjcol

%acronyms
\acrodef{FC}{Fully Connected}
\acrodef{AI}{Artificial Intelligence}
\acrodef{GELID}{GamEpLay issue detector}
\acrodefplural{GUI}[GUIs]{graphical user interfaces}
\acrodef{GUI}{graphical user interface}
\acrodef{DRL}{deep reinforcement learning}
\acrodef{RL}{reinforcement learning}
\acrodef{WOB}{World of Bugs}
\acrodef{ABD}{Automated Bug Detection}
\acrodef{JSON}{JavaScript Object Notation}
\acrodef{DBSCAN}{Density-Based Spatial Clustering of Applications with Noise}
\acrodef{LOD}{Level of Detail}
\acrodef{CFT}{Hybrid Co-Fine Tuning}
\acrodef{JEPA}{Joint Embedding Predictive Architecture}
\acrodef{ViT}{Vision Transformer}
\acrodef{fps}{frames per second}
\acrodef{ConvLSTM}{Convolutional Long Short-Term Memory}
\acrodef{RCNN}{Region-based Convolutional Neural Network}
\acrodef{BRNN}{Bidirectional Recurrent Neural Network}
\acrodef{LSTM}{Long Short-Term Memory Network}
\acrodef{ResNet-BiLSTM}{ResNet and Bidirectional Long Short-Term Memory}
\acrodef{BiLSTM}{Bidirectional LSTM}
\acrodef{ASL}{Asymmetric Loss}
\acrodef{I3D}{Inflated 3D ConvNet}
\acrodef{R3D-18}{3D ResNet}
\acrodef{GRU}{Gated Recurrent Unit}
\acrodef{MR}{Mixed Reality}
\acrodef{AR}{Augmented Reality}
\acrodef{VR}{Virtual Reality}
\acrodef{3DoF}{Three Degrees of Freedom}
\begin{abstract}
Recent years have seen an increasing need for \ac{MR} tools for game development and visual game analytics. However, the number of visual game analytics tools for \ac{MR} games remains scarce due to the unique complexities in integrating real components from both the physical and the digital worlds in \ac{MR} games. Existing visual analytics tools for such games are not suitable for space-agnostic generalization. To tackle this problem, we introduce a space-agnostic visual game analytics tool that incorporates adaptive spatial reconstruction and real-time object detection built for the Microsoft HoloLens 2 \ac{MR} headset. We believe that the outcomes of our work will increase insights into player environment and pave a new direction for visual game analytics for \ac{MR} game development.
\end{abstract}


\section{Introduction}
The recent rapid advancements in \ac{MR} have introduced an opportunity for new visual game analytics tools. However, \ac{MR} being distinct from other gaming platforms comes with its unique complexities for the integration of physical and digital worlds. This, in turn, creates a unique gaming experience for which a different approach to game analytics becomes essential. Integrating real elements from the physical world and virtual elements from the digital world creates a new kind of complexity for player interactions.

\ac{MR}  devices such as the Microsoft HoloLens 2 face several hardware limitations, including the lack of a wide field of view \cite{mivitals} and imperfect hand tracking \cite{tripad}, which are essential for a positive experience for the user. Therefore, game developers and designers must address these issues by analyzing gameplay data within the \ac{MR} environment using \ac{MR} hardware to gain deeper contextual information about game dynamics and player behaviour. This, in turn, can lead to a more immersive gaming experience.

The highly immersive nature of \ac{MR} is key to understanding player behaviour in a more natural gaming environment. \ac{MR} devices allow developers to conduct real-time analysis of player behaviour in 3D surroundings with insights into how players navigate the environment and 
respond to stimuli. This approach can potentially enhance immersion and engagement. \ac{MR} also allows intuitive exploration and manipulation of data in 3D. Enhancing the understanding of complex datasets is also known as immersive analysis, which has significant potential \cite{immersive, marriott2018immersive, lee2022design}. \ac{MR}, among others, provides data on player movement in physical space and hand tracking. These new multimodal insights permit the analysis of visual, auditory, and haptic information, which can be used for a better understanding of gameplay, as well as of physical and cognitive load. 

The existing landscape of game analytics tools leaves a gap for developers working on \ac{MR} games. Despite significant progress in immersive analytics, there is still a major gap in managing complex and precise input data that hinders visual analysis for \ac{MR} gaming environments \cite{zhang2024exploring} along with the absence of tools that can reconstruct gameplay sessions without being dependent on a specific location. Because of this problem, game developers may end up resorting to labour-intensive and time-consuming practices of manually reviewing lengthy video recordings of player game sessions in a specific location to detect bugs and to address them.

To address this gap, we propose a space-agnostic game analytics tool for \ac{MR} games that offers a toolset for the analysis and evaluation of player behaviour, which extends our previous work \cite{gamr}. We make the following contributions:

\begin{itemize}
    \item a space-agnostic visual analytics tool for \ac{MR} games with analytical capabilities, including reconstructed gameplay sessions, data visualization capabilities, and an object-tracking system;

    \item a spatial reconstruction system to reconstruct gameplay environment independent of the location of the gameplay session.
\end{itemize}

\section{Related Work}
None of the existing work has provided space-agnostic spatial reconstruction support for visual analytics tools on \ac{MR} platforms, as current approaches typically depend on fixed environments to perform visual analysis of gameplay sessions.
\subsection{Game Analytics}
Medler et al. \cite{medler2011data} developed Data Cracker that uses telemetric "hooks" within the Dead Space 2 game code repository to gather player data to be visually aggregated later, but our tool collects data independently from the code, simplifying integration. Stahlke et al. \cite{stahlke2020artificial} developed \textit{PathOS} that utilizes a model of player perception, memory, and cognitive architecture to simulate visual and information processing through individual paths of agents mimicking player navigation, as well as heatmaps and bubble visualizations. MacCormick and Zaman \cite{maccormick2020echo}  analyze player behaviour by reconstructing gameplay from recorded data and replaying it from different perspectives. Sargolzaei et al. \cite{gamr} developed an analytics tool for \ac{MR} games that supports developers in identifying and addressing gameplay issues through reconstructed sessions, heatmaps, annotations, and multimodal tracking of hands, camera, input, and audio.
\subsection{Spatio-temporal Analytics}
Kang and Kim \cite{kang2015automated} examined spatio-temporal analysis with visualization approaches, trajectories, and an in-house telemetry system to provide an understanding of game data. They proposed how to visualize player behaviour in the game world using spatio-temporal data directly. The spatio-temporal analysis integration is available in our tool.  However, their work is dependent on fixed environments, unlike in our tool, for which the visual analysis of a gameplay session is not dependent on fixed locations or environments.  Wallner et al. \cite{wallner2012spatiotemporal} created a tool that visualizes the game space using nodes representing areas players traverse to provide a holistic view of player movement and interactions, enabling users to discern player behaviour, patterns, game design and flow. Kloiber et al. \cite{kloiber2020immersive} visualize human motion, analyzing behaviour through motion paths and hand movements. Similarly, our tool employs hand motion tracking and player trajectory trails using coloured trails to study player movement and actions within the game.

\subsection{Temporal Analytics}
A framework called TRUE \cite{jun} was developed by Kim et al. to capture analytics for intricate systems. In this framework, players can gain feedback from events recorded with timestamps and integrated questionnaires. Although both our tool and TRUE aim to support comprehensive data capture and analysis, our approach focuses on space-agnostic spatial reconstruction for data collection while also enriching the captured data through real-time object recognition during playback, enabling additional information about environmental objects to be identified and analyzed. GameVis \cite{feitosa} was developed by Feitosa et al., aiding developers in creating tailored analytics visualizations by incorporating game data visualization into web technologies using their framework, which emphasizes the importance of time-based representation.
\begin{figure}[h!]
    \centering
    \includegraphics[width=\columnwidth, height=0.6\textheight, keepaspectratio]{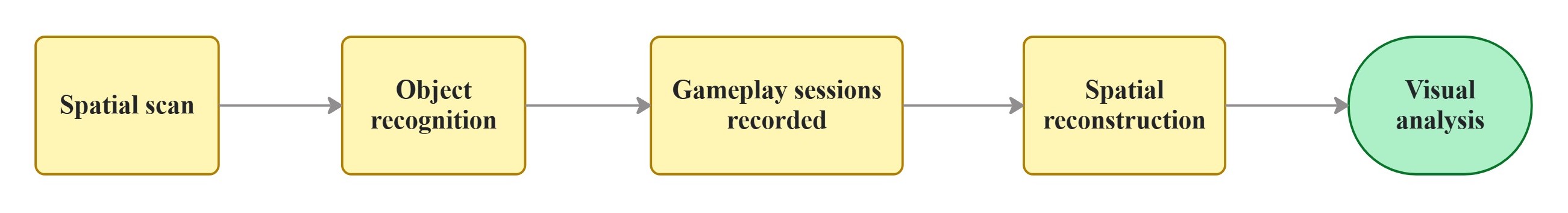}
    \caption{The pipeline in GAMR+}
    \label{fig:gamrplus-pip}
\end{figure}
\subsection{Mixed Reality Analytics}
Sicat et al. \cite{sicat2018dxr} introduced their Unity toolkit called \textit{DXR} for extended reality (XR) applications in the Unity engine that allows users to create immersive 3D visualizations without needing expertise in 3D graphics, similar to that of our tool. MIRIA, a toolkit developed by Büschel et al. \cite{buschel2021miria}, allows users to craft in-situ visualizations and analyze interactions within \ac{MR} environments with a range of visualizations, including heatmaps, 3D trajectories, and scatterplots for holistic data analysis for spatial data in \ac{AR} and \ac{MR} applications, but not for games. Similarly, our tool provides in-situ data visualizations for MR games, allowing insights into player interactions. AvatAR developed by Reipschläger et al. \cite{reipschlager2022avatar}, visualizes human motion using 3D trajectories and virtual avatars with a detailed representation of movement and posture. Hubenschmid et al. \cite{hubenschmid2022relive} developed ReLive for analyzing \ac{MR} user studies by interacting with recorded spatial data within the original environment. Simonson et al. \cite{depth} developed \textit{DEPTH}  that offers a voxel-based visualization to automatically pinpoint spatial design flaws in \ac{VR} training simulations.

\begin{figure*}[t]
    \centering
    \begin{subfigure}[b]{0.3\textwidth}
        \includegraphics[width=\linewidth]{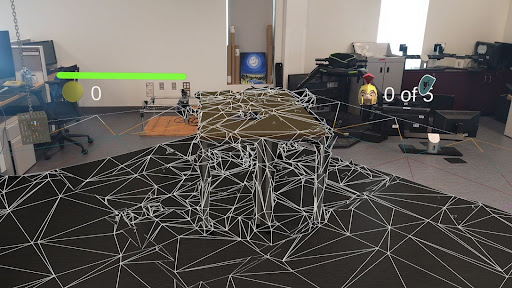}
        \caption{Spatial scan}
        \label{fig:spatial-scan}
    \end{subfigure}
    \hfill
    \begin{subfigure}[b]{0.3\textwidth}
        \includegraphics[width=\linewidth]{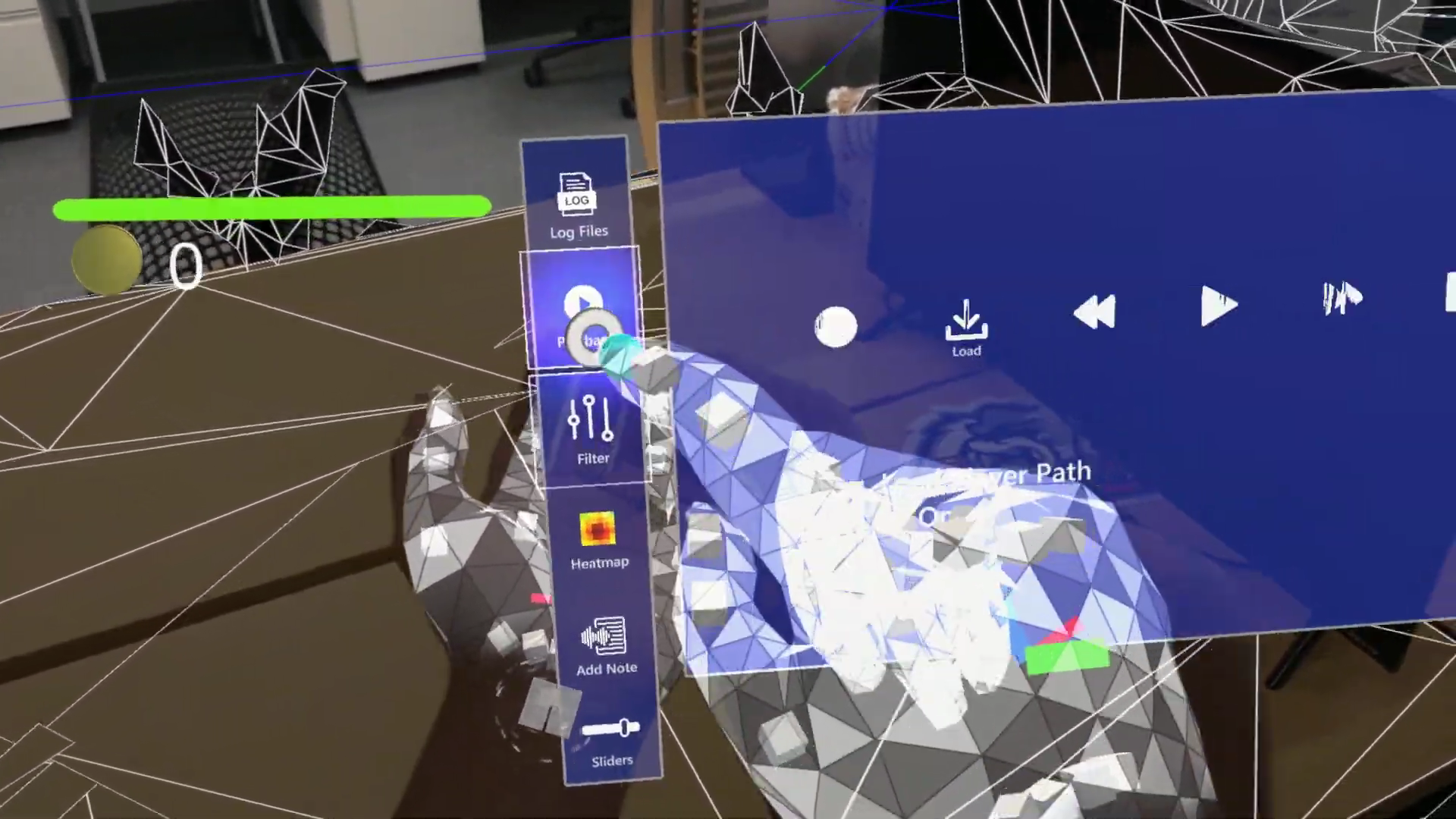}
        \caption{Recording and replay UI}
        \label{fig:ui}
    \end{subfigure}
    \hfill
    \begin{subfigure}[b]{0.3\textwidth}
        \includegraphics[width=\linewidth]{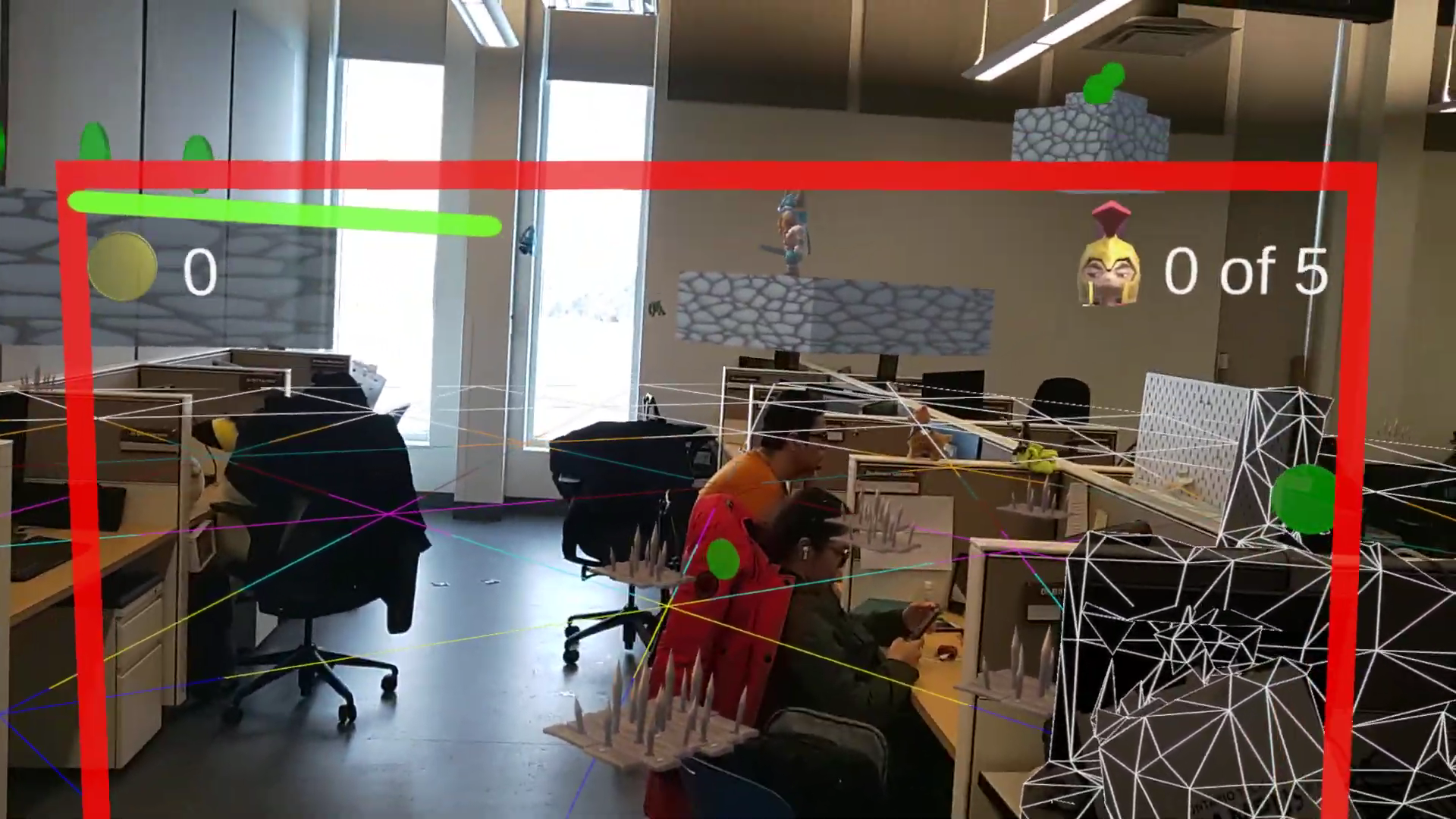}
        \caption{Recording gameplay}
        \label{fig:record}
    \end{subfigure}

    \vspace{0.5em}

    \begin{subfigure}[b]{0.3\textwidth}
        \includegraphics[width=\linewidth]{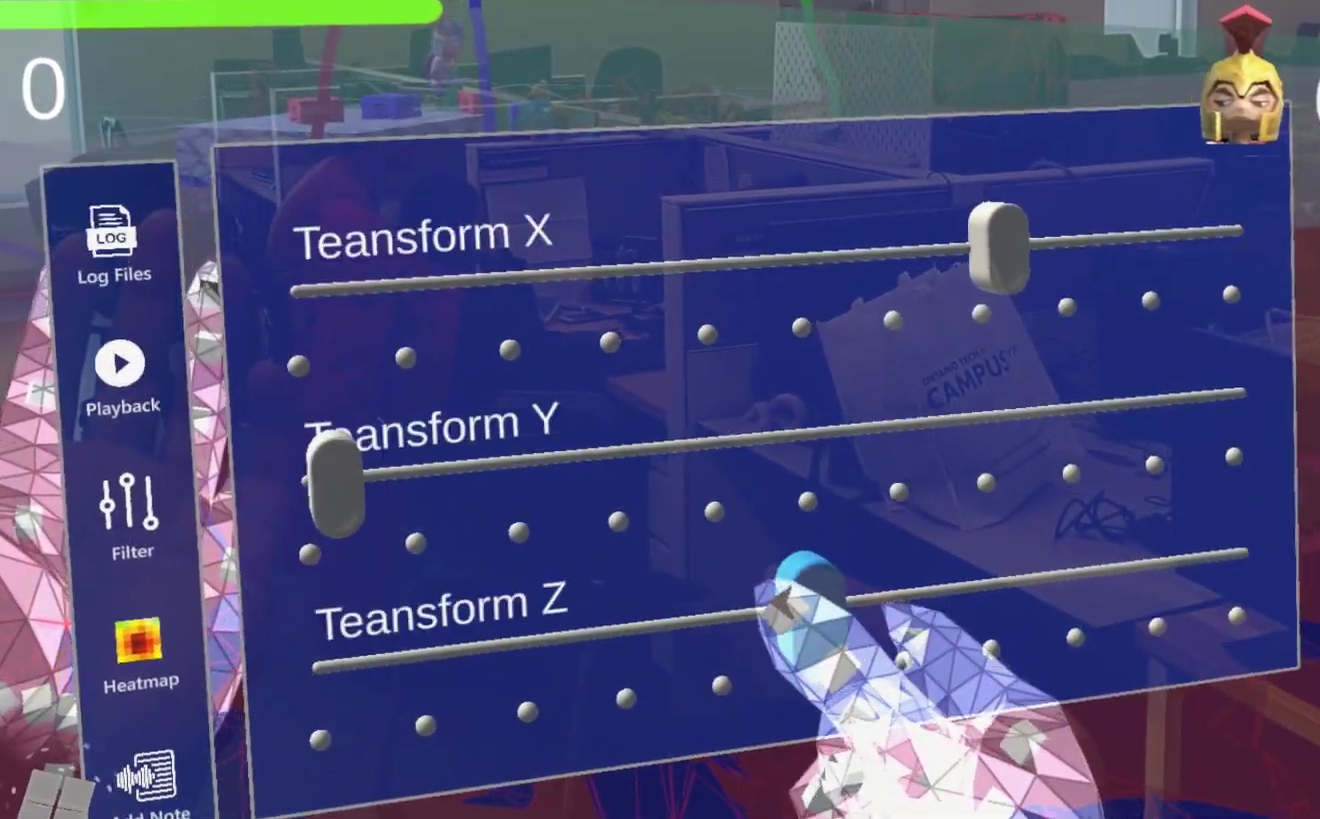}
        \caption{Mesh position controls}
        \label{fig:sliderui}
    \end{subfigure}
    \hfill
    \begin{subfigure}[b]{0.3\textwidth}
        \includegraphics[width=\linewidth]{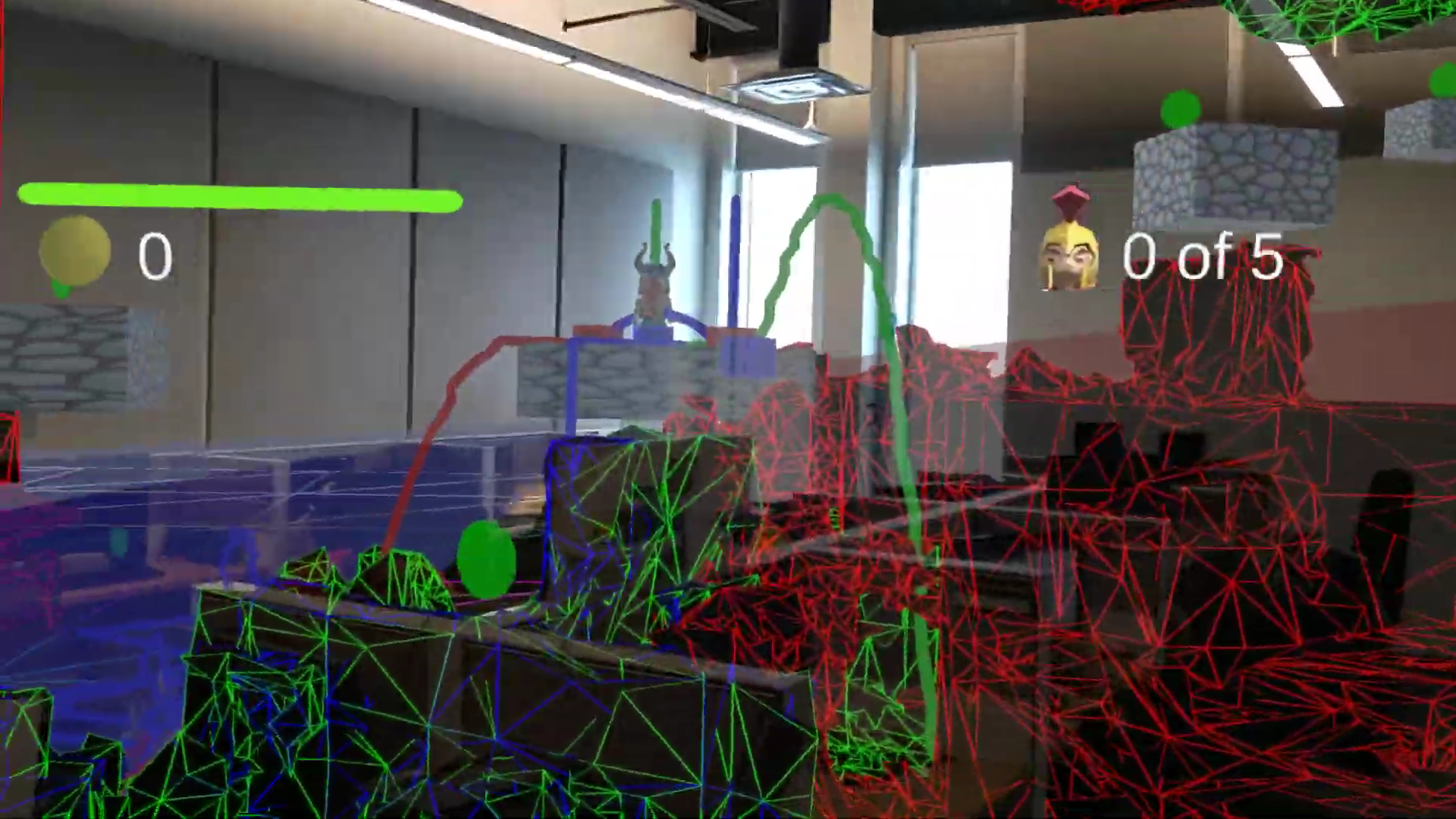}
        \caption{Spatial reconstruction from sessions}
        \label{fig:replay}
    \end{subfigure}
    \hfill
    \begin{subfigure}[b]{0.3\textwidth}
        \includegraphics[width=\linewidth]{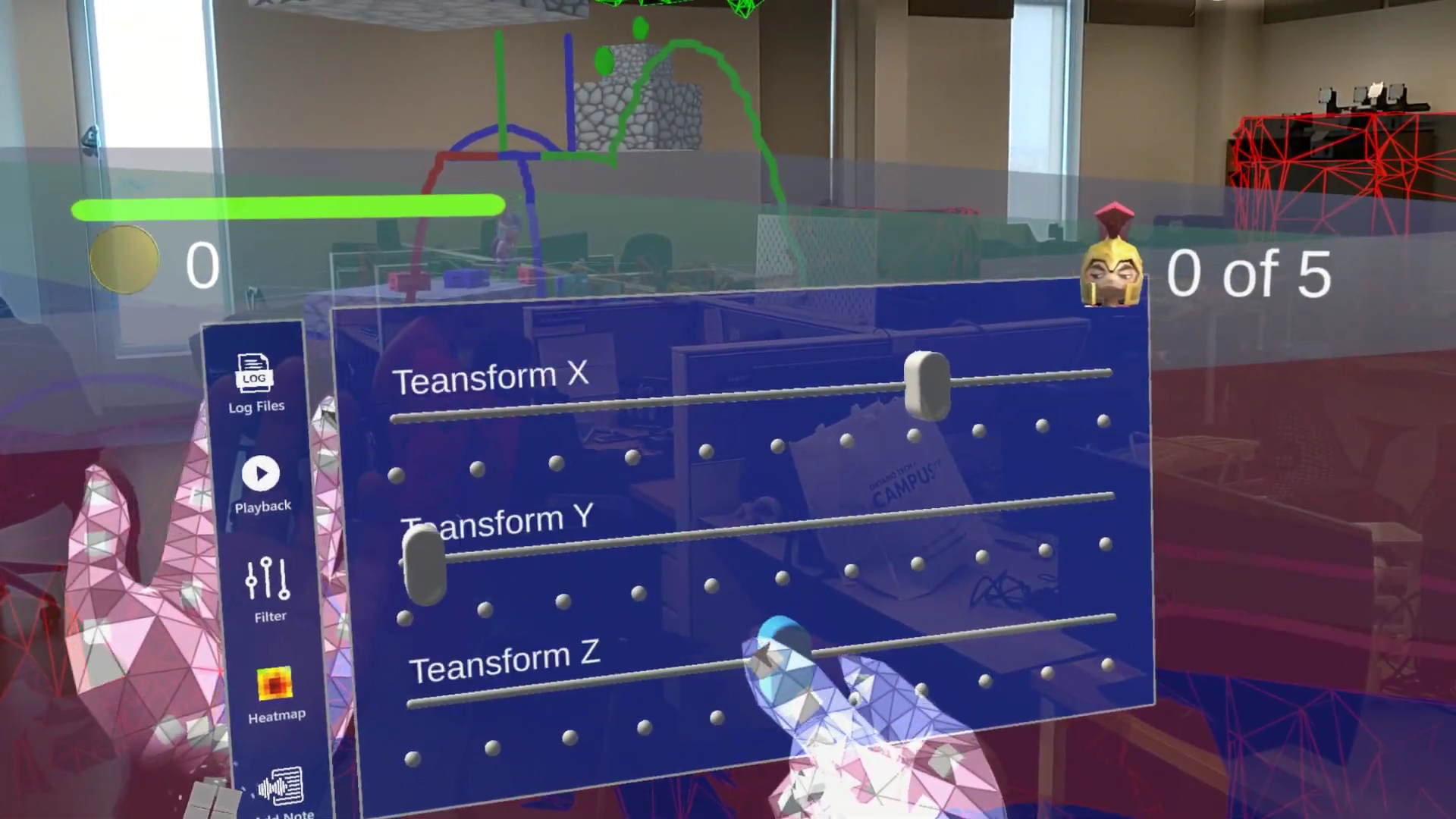}
        \caption{Moving reconstructed meshes}
        \label{fig:slider2}
    \end{subfigure}

    \caption{GAMR+ UI and usage}
    \label{fig:gamrplus}
\end{figure*}
\section{GAMR+}
Our visual analytics tool for \ac{MR} games, GAMR+, directly extends GAMR \cite{gamr}. GAMR+ reconstructs recorded gameplay sessions from the Microsoft HoloLens 2 \ac{MR} headset\footnote{\url{https://learn.microsoft.com/en-us/hololens/}}, enabling game developers to analyze player behaviour and the surrounding environment regardless of where the gameplay session took place. It consists of two components, recording and visualization, and includes adaptive spatial reconstruction and object recognition systems. The aforementioned features of GAMR+ were introduced keeping the intricacies of \ac{MR} games in mind. The pipeline (Fig. \ref{fig:gamrplus-pip}) starts off with scanning the game environment by firing rays to capture necessary data for mesh creation and recognizing objects in the space. Afterwards, the user can play the game and record multiple gameplay sessions. After recording gameplay sessions, the user can reconstruct the game environment in another location from the original location of the gameplay recording using our adaptive spatial reconstruction system, which is space-agnostic. Finally, the user can visualize and compare multiple recorded gameplay sessions in the reconstructed environment. See the UI in Fig.~\ref{fig:gamrplus}.

\subsection{Adaptive Spatial Reconstruction}
GAMR+ features a space-agnostic adaptive spatial reconstruction system. This tool is built with the spatial awareness API of the Mixed Reality Toolkit 2 (MRTK 2) for the Microsoft HoloLens 2 \ac{MR} headset on the Unity game engine. Firstly, the headset's cameras and sensors scan the gameplay environment and save the environment data in the database when the game is initially opened (Fig. ~\ref{fig:spatial-scan}). The start of the gameplay recording (Fig.~\ref{fig:ui}) is indicated by the red bounding box (Fig. ~\ref{fig:record}). 
When recording is finished, the user can load the game environment and meshes anywhere (Fig. ~\ref{fig:replay}) to view the player’s path, multiple gameplay sessions, and environments, each indicated using different colours.
 It is particularly useful since the user can change locations but still have access to the same environment of the recorded gameplay sessions depicted by the reconstructed meshes. In the tool, the UI has a slider tab (Fig. ~\ref{fig:sliderui}), which has three sliders to control the position of the reconstructed meshes. When the reconstructed meshes of saved gameplay environments are loaded and displayed in different colours (Fig.~\ref{fig:replay}), the user can open the slider tab while the meshes are visible. Afterwards, the user can use the sliders as shown in Fig.~\ref{fig:slider2} to move all the meshes with \ac{3DoF}.

\begin{figure}[h!]
    \centering
    \includegraphics[width=1\columnwidth, height=0.6\textheight, keepaspectratio]{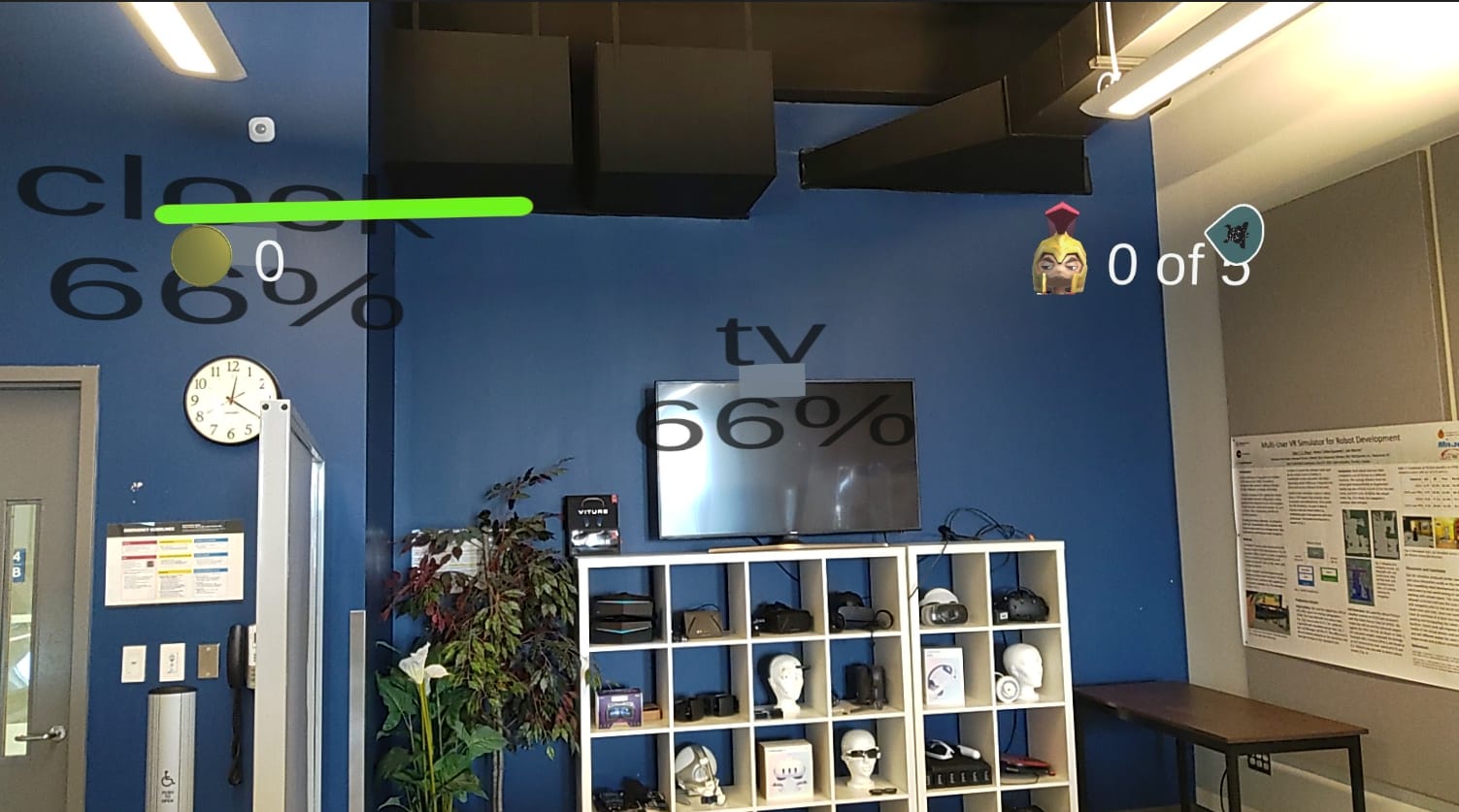}
    \caption{Real-time object recognition with confidence scores of detected objects}
    \label{fig:object}
\end{figure}
\subsection{Real-Time Object Detection}
When GAMR+ is launched, it performs real-time object recognition from the HoloLens headset's camera feed. To implement this, we used the Unity Barracuda library and a lightweight version of YOLOv8\footnote{\url{https://github.com/ultralytics/ultralytics}}  with the use of the ONNX format. The lightweight YOLOv8 model can detect 80 different classes of objects, including different kinds of furniture, accessories, indoor items, electronics, appliances, kitchenware, etc., with good confidence scores. The model is deployed on the headset and integrated into GAMR+. Fig.~\ref{fig:object} shows real-time object detection in progress in an Action-Adventure game environment before gameplay. Object detection during the initial spatial scan can be integrated into the spatial mesh data in future iterations of our tool, which can be used to get better insights into the objects that were present in the gameplay session labelled with the reconstructed mesh. This may help users to get insights into the game character's interaction with different objects in its environment, potentially helping users to detect issues in the interactions.
\vspace{-6 pt}
\subsection{Action-Adventure game}
\begin{figure}[h!]
    \centering
    \includegraphics[width=1\columnwidth, height=0.6\textheight, keepaspectratio]{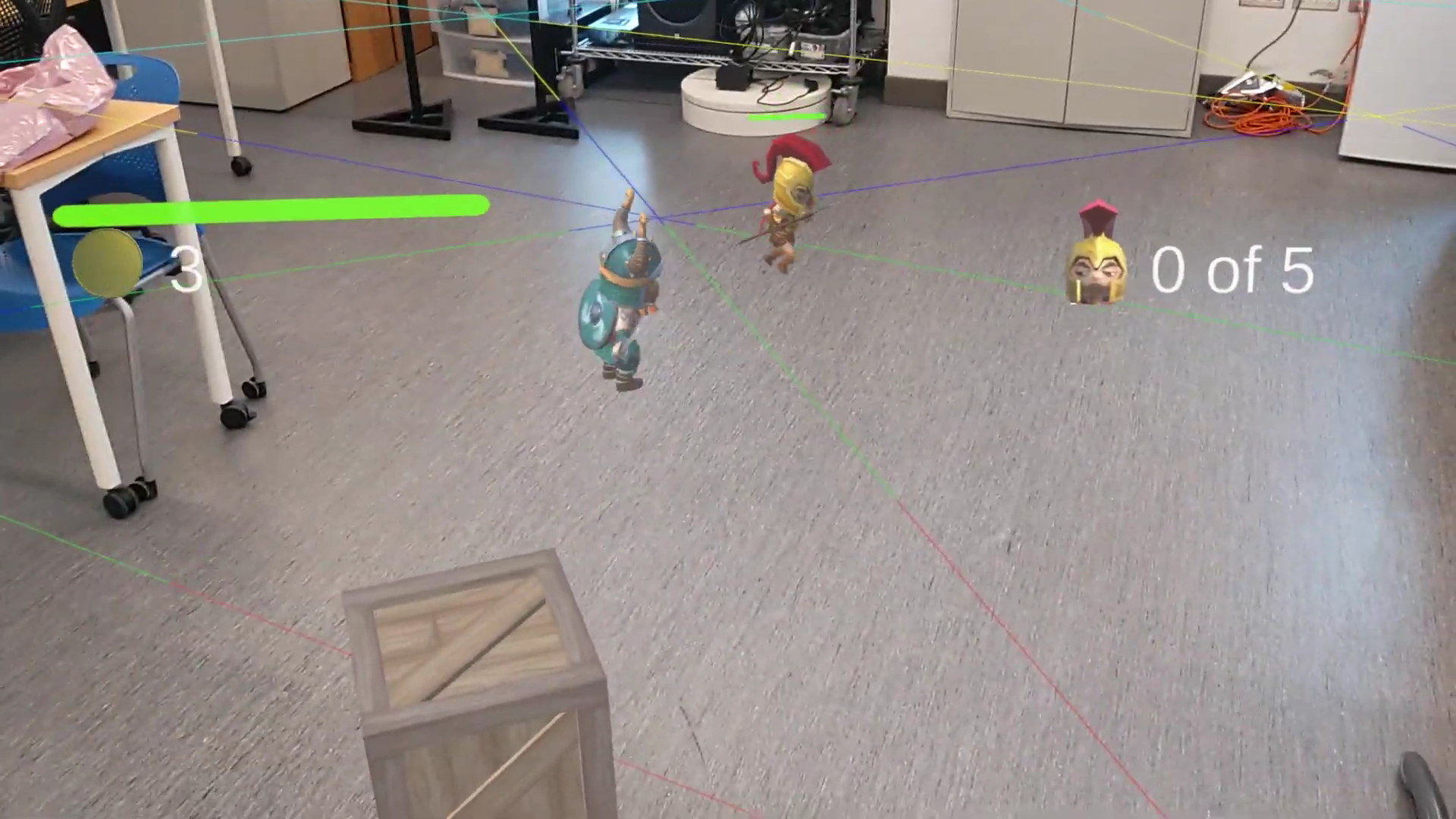}
    \caption{Action-Adventure game}
    \label{fig:game}
\end{figure}

We tested GAMR+ in the Action-Adventure game \cite{gamr} where the player-controlled character wears green armour and is controlled using an Xbox controller after a spatial scan and object recognition of the environment. The character's health bar appears in light green in the top-left corner (Fig. ~\ref{fig:game}). The objective of the game is to defeat enemy characters dressed in golden armour. The game ends when the boss is defeated. The boss battle is activated after a number of enemies in the game have been defeated (five in this example).
\section{Discussion}
GAMR+ was developed with the goal of creating a versatile and seamless visual analytics tool for \ac{MR} gameplay analysis. It offers a distinct approach that integrates real-time object detection with adaptive spatial reconstruction of the gameplay environment, supporting location-independent analysis of gameplay sessions, including those recorded elsewhere. GAMR+ is designed to complement traditional evaluation processes, such as video footage review and automated bug detection. It presents a system that provides a space-agnostic gameplay environment and session reconstruction for extending the modalities of data collection in a complex environment like an \ac{MR} game. 

This work has some notable limitations affecting its validity and generalizability. Firstly, the current iteration of the tool has only been tested on a single in-house built game. This restriction limits generalization across different games and genres. Secondly, the limitations of the Microsoft HoloLens 2 headset due to its lack of computational power limits the ability to deploy the state-of-the-art object detection models on the tool. This causes delay and some inaccuracy in real-time object detection. Thirdly, spatial analytics used in GAMR+ can cause ethical and privacy risks to users. Lastly, the current iteration of the tool lacks a user study to evaluate the tool with \ac{MR} game developers. 

\section{Conclusion and Future Work}
In this work, we present a work-in-progress tool for space-agnostic visual analytics in \ac{MR}. The tool aims to provide game developers with better insights into the \ac{MR} game environment to detect bugs and to address them accordingly. We introduced adaptive spatial reconstruction to make the reconstruction of gameplay sessions independent of the location of the original gameplay footage recording. As a proof of concept, we tested the functionality of the tool on a game that was in-house developed. 

For future directions of our work, GAMR+ can be deployed as a separate application that can run on any \ac{MR} headset and games that allow spatial meshing capabilities, which would make the tool cross-platform and game-agnostic through game engine plugins. Secondly, we aim to move the object detection models to cloud servers to address the problem of computational limitations of the headset hardware, which would solve the problem of deploying larger and more accurate object detection models onto the headset. Lastly, we plan to evaluate the tool in a user study involving with \ac{MR} game developers to study its effectiveness.

\section*{Acknowledgment}

This research was supported by NSERC DG and Mitacs.
% The preferred spelling of the word ``acknowledgment'' in America is without 
% an ``e'' after the ``g''. Avoid the stilted expression ``one of us (R. B. 
% G.) thanks $\ldots$''. Instead, try ``R. B. G. thanks$\ldots$''. Put sponsor 
% acknowledgments in the unnumbered footnote on the first page.

\bibliographystyle{IEEEtran}
\bibliography{references}
% \vspace{12pt}

\end{document}